\documentclass[12pt]{article}
\usepackage{amsmath,amssymb,amsfonts}
\usepackage[paper=letterpaper,margin=1.0in]{geometry}
\usepackage{graphicx}

\newcommand{\be}{\begin{equation}}
\newcommand{\ee}{\end{equation}}
\newcommand{\bea}{\begin{eqnarray}}
\newcommand{\eea}{\end{eqnarray}}
\newcommand{\ba}{\begin{array}}
\newcommand{\ea}{\end{array}}
\newcommand{\ben}{\begin{enumerate}}
\newcommand{\een}{\end{enumerate}}
\newcommand{\bi}{\begin{itemize}}
\newcommand{\ei}{\end{itemize}}
\newcommand{\bc}{\begin{center}}
\newcommand{\ec}{\end{center}}
\newcommand{\bfig}{\begin{figure}}
\newcommand{\efig}{\end{figure}}
\newcommand{\bq}{\begin{quotation}}
\newcommand{\eq}{\end{quotation}}
\newcommand{\bt}{\begin{table}}
\newcommand{\et}{\end{table}}
\newcommand{\btab}{\begin{tabular}}
\newcommand{\etab}{\end{tabular}}
\newcommand{\bs}{\begin{slide}}
\newcommand{\es}{\end{slide}}

\begin{document}

{\footnotesize
${}$
}

\bc

\vskip 1.0cm
\centerline{\Large \bf Quantum Gravity and Dark Matter}
\vskip 0.5cm
\vskip 1.0cm

\renewcommand{\thefootnote}{\fnsymbol{footnote}}

\centerline{{\bf
Chiu Man Ho${}^{1}$\footnote{\tt chiuman.ho@vanderbilt.edu},
Djordje Minic${}^{2}$\footnote{\tt dminic@vt.edu}, and
Y.\ Jack Ng${}^{3}$\footnote{\tt yjng@physics.unc.edu}
}}

\vskip 0.5cm

{\it
${}^1$Department of Physics and Astronomy, Vanderbilt University,\\
Nashville, TN 37235, U.S.A.\\
${}$ \\
${}^2$Department
of Physics, Virginia Tech, Blacksburg, VA 24061, U.S.A. \\
${}$ \\
${}^3$Institute of Field Physics, Department of Physics and
Astronomy,\\
University of North Carolina, Chapel Hill, NC 27599, U.S.A.
}

\vspace{3cm}

``This paper has been awarded the Fifth Prize in the 2011 Essay Competition of the Gravity Research Foundation'' \\

\ec

\vskip 1.0cm

\begin{abstract}

We propose a connection between global physics and local galactic
dynamics via quantum gravity.  The salient features of cold dark matter (CDM)
and modified Newtonian dynamics (MOND) are combined into a unified scheme by
introducing the concept of MONDian dark matter which behaves like CDM at
cluster and cosmological scales but emulates MOND at the galactic scale.

\end{abstract}

\renewcommand{\thefootnote}{\arabic{footnote}}

\newpage

\section{Introduction}

The vexing problem of "missing mass" has been with us ever since
Zwicky found that the individual galaxies in the Coma cluster were
zipping around too fast for gravity to hold them together in a cluster.
This problem started to get the attention of the scientific community
after Rubin and her collaborators discovered that the clouds of
hydrogen gas in several distant galaxies were orbiting the center of
their galaxies at a speed far exceeding what could be accounted for by
the gravitational pull due to the mass of visible baryonic matter.  This
mass mismatch has since been astoundingly confirmed from dwarf galaxies
to galaxy groups.  A simple but bold solution would be to postulate the
existence of dark (invisible) matter that makes up the mass difference
\cite{dark}.

This apparent need for dark matter at the galactic scale is even more
urgent at larger scales.  Dark matter is needed to account for the
correct cosmic microwave background spectrum shapes (including the
alternating peaks).  It is also needed to yield the correct large-scale
structures and elemental abundances from big bang nucleosynthesis.
(Supercomputer simulations suggest that the cosmos
would look very different if dark matter did not exist.)
Dark matter also provides the correct
gravitational lensing.  One of the most prominent examples pointing to
the existence of dark matter is the Bullet Cluster, a pair of merging
galaxy clusters.  In this system, the gravitational lensing of
background galaxies indicates that the mass is offset from the X-ray
plasma, suggesting that dark matter has shifted the center of gravity
elsewhere.  The consensus in the cosmology community appears to be that
dark matter exists!  By now all this has been canonized in the concordant
$\Lambda$CDM model of cosmology according to which cold dark matter
accounts for about $23 \%$ of the energy and mass of the universe,
$5 \%$ resides in ordinary matter, and dark energy in the form of
cosmological constant has the lion's share, accounting for the remaining
$72 \%$ (numerically the cosmological constant is related to the current
Hubble parameter as $\Lambda \sim 3 H^2$.)

But there is a fly in the ointment.  At the galactic scale, dark matter
can explain the observed asymptotic independence of orbital velocities
on the size of the orbit only by fitting data (usually with two
parameters) for individual galaxies.  It can do no better in explaining
the observed baryonic Tully-Fisher relation \cite{TF,McGaugh}, i.e., the
asymptotic-velocity-mass ($v^4 \propto M$) relation.  Another problem
with dark matter is that it seems to possess too much power on small
scales ($\sim 1 - 1000$ kpc) \cite{cen}.  While cold dark matter works spectacularly
well at the cluster and cosmic scales, it had been found to be somewhat
wanting at the galactic scale.

This is in stark contrast to another paradigm that goes by the name of
modified Newtonian dyanmics (MOND) \cite{mond,teves},
due to Milgrom, which stipulates that the
acceleration of a test mass $m$ due to the source $M$ is given by
$a= a_N$ for $a \gg a_c$, but
$a = \sqrt{a_N\, a_c}$ for $a \ll a_c$,
where $a_N= G M /r^2$ is the magnitude of
the usual Newtonian acceleration and the critical acceleration
$a_c$ is numerically related to the speed of light $c$ and
the Hubble scale $H$ as
$a_c \approx c H/(2 \pi) \sim 10^{-8} cm/s^2.$
With only a single parameter
MOND can explain easily and rather successfully the observed
flat galactic rotation curves and the observed Tully-Fisher relation
\cite{dsmond}.
But there are problems with MOND at the cluster and cosmological
scales.  The reason may be due to the lack of a fundamental
relativistic theory of MOND.

Obviously it is desirable to combine the salient successful features
of both CDM and MOND into a unified scheme.  This essay describes our
effort in that endeavor.  Succinctly, by making use of a novel
quantum gravitational interpretation of (dark) matter's inertia
our scheme \cite{HMN} indicates that dark matter emulates MOND at the
galactic
scale.  This work embraces a recurring theme at the current gravity
research frontiers: the relationship between gravity and thermodynamics.

\section{Entropic Gravity and Critical Galactic Acceleration}

We start with
the recent work of E. Verlinde \cite{verlinde,Jacob95,Pad,Smolin}
in which the canonical Newton's laws are derived
from the point of view of holography.
Using the first law of thermodynamics, Verlinde proposes the concept of
entropic force
$
F_{entropic} = T \frac{\Delta S}{\Delta x},
$
where $\Delta x$ denotes an infinitesimal spatial displacement of a
particle with mass $m$ from the heat
bath with temperature $T$. He then
invokes Bekenstein's original arguments
concerning the entropy $S$ of black holes \cite{bekenstein}
by imposing $
\Delta S = 2\pi k_B \frac{mc}{\hbar} \Delta x
$.\,
Using the famous formula for the Unruh temperature,
$
k_B T = \frac{\hbar a}{ 2\pi c},
$\,
associated with a uniformly accelerating (Rindler) observer
\cite{unruh,Davies},
he obtains
$F_{entropic}= T \nabla_x S= m a$,
Newton's second law (with the vectorial form
$
\vec{F} = m \vec{a},
$\,
being dictated by the gradient of the entropy).

Next, Verlinde considers an
imaginary quasi-local (spherical) holographic screen of area $A=4 \pi
r^2$ with
temperature $T$. Then, he assumes the equipartition of energy $E=
\frac{1}{2} N k_B T$ with $N$ being
the total number of degrees of freedom (bits) on the screen given by $N =
Ac^3/(G \hbar)$. Using the Unruh
temperature formula and the fact that $E=M c^2$, he obtains
$
2 \pi k_B T = G M /r^2
$
and recovers exactly the non-relativistic Newton's law
of gravity, namely $a= G M /r^2$.  But this is precisely the
fundamental relation that Milgrom
is proposing to modify so as to fit the galactic rotation curves.
Therefore it is now natural to ask whether there is an
entropic \cite{bekenstein} or
holographic \cite{hawking,holography,Susskind,adscft}
interpretation behind Milgrom's modification of Newton's second law.

We first have to recognize that we live in an accelerating universe (in
accordance with the $\Lambda$CDM model).
This suggests that we will need a generalization \cite{HMN} of Verlinde's
proposal \cite{verlinde}
to de Sitter space with a positive cosmological constant (which,
we recall, is
related to the Hubble parameter $H$ by $\Lambda \sim 3 H^2$).
The Unruh-Hawking temperature
as measured by a non-inertial observer with acceleration $a$ in the de
Sitter space is given by
$ \sqrt{a^2+a_0^2}/(2 \pi k_B)$\, \cite{deser,Jacob98}, where
$a_0=\sqrt{\Lambda/3}
$ \,\cite{hawking}.
Consequently, we can define the net temperature measured by the
non-inertial observer (relative to the inertial observer)
to be $\tilde{T} = [(a^2+a_0^2)^{1/2} - a_0]/(2 \pi k_B)$.

We can now follow Verlinde's approach.\footnote{
We only need to replace the $T$ in Verlinde's argument
by $\tilde{T}$ for the Unruh
temperature.}  Then the entropic force, acting on the test mass $m$ with
acceleration $a$ in de Sitter space, is given by
$F_{entropic}=\tilde{T}\, \nabla_x S= m [(a^2+a_0^2)^{1/2}-a_0].$
For $a \gg a_0$, the entropic force is given by $F_{entropic}\approx
ma$, which gives $ a = a_N$ for a test mass $m$ due to the source $M$.
But for $a \ll a_0$, we have
$F_{entropic}\approx ma^2/(2a_0)$ which, upon equated with $m a_N$,
yields $a \approx \sqrt{2 a_N\, a_0}$.  Thus we have derived Milgrom's
scaling or MOND with the correct order of magnitude for the (observed)
critical galactic acceleration
$a_c = 2 a_0 \sim \sqrt{\Lambda/3} \sim H \sim 10^{-8} cm/s^2$!
\footnote{Actually
Milgrom did observe \cite{interpol} that the generalized Unruh
temperature
$\tilde{T}$ can give the correct behaviors of the interpolating function
between the usual Newtonian
acceleration and his suggested MONDian deformation for very small
accelerations.  He was right; but, unlike us, he could offer no
justification.}
From our perspective, MOND {\it is a (successful) phenomenological
consequence of quantum gravity}.

Having derived MOND we can now write the entropic force, in the
regime $a \ll a_0$, as
$F_{entropic} \approx m \frac{a^2}{2\,a_0} = F_{Milgrom} \approx m
\sqrt{a_N a_c}$ implying that
$ a = \left(\,4\, a_N \,a_0^2 \,a_c \,
\right)^{\frac14}
=\left(\,2\, a_N \,a_0^3 / \pi \, \right)^{\frac14}$.
Numerically, it turns out that $2 \pi a_c \approx a_0$, and so we set
$a_c = a_0/(2\pi )$ for simplicity.
The flat galactic rotation curves as well as the Tully-Fisher relation
can be easily recovered by using
$
F_{centripetal} = m \frac{a^2}{2\,a_0} = \frac{m \,v^2}{r}
$
to solve for the terminal velocity $v$.

\section{MONDian Dark Matter}

To see how dark
matter can behave like MOND at the galactic scale,
let us continue to follow Verlinde's holographic approach. Invoking the
imaginary holographic screen of radius $r$, we can write\footnote{
We need to replace the $T$ and $M$ in Verlinde's
argument by $\tilde{T}$ and $\tilde{M}$ respectively.}
$2 \pi k_B \tilde{T} = \frac{G\,\tilde{M}}{r^2}$,
where $\tilde{M}$ represents the \emph{total} mass enclosed within the
volume $V = 4 \pi r^3 / 3$. But, as we will show below, consistency with
the discussion in the previous section (and with observational data)
demands that
$\tilde{M} = M + M'$ where $M'$ is some unknown mass --- that is, dark
matter. Thus, we need the concept of dark matter for consistency.

First note that it is natural to write the entropic force
$F_{entropic} = m [(a^2+a_0^2)^{1/2}-a_0]$ as
$F_{entropic} = m\,a_N [1 +  (a_0/a)^2/ \pi]$ since the latter expression
is arguably the simplest interpolating formula
for $F_{entropic}$ that satisfies the two requirements: $a \approx (2 a_N
a_0^3/ \pi)^{1/4}$ in the small acceleration $a \ll a_0$
regime, and $a = a_N$ in the $a \gg a_0$ regime.
But we can also write $F$ in another, yet equivalent, form:
$F_{entropic} = mG(M+M')/r^2$.
These two forms of $F$ illustrate the idea of CDM-MOND duality \cite{HMN}.
The first form can be interpreted to mean that there is no dark matter,
but that the law of gravity is modified, while the second form means
that there is dark matter (which, by construction, is consistent with
MOND) but that the law of gravity is not modified.
The second form gives us a very intriguing
dark matter profile: $M'=\frac{1}{\pi} \,\left(\,\frac{a_0}{a}\,\right)^2
\, M$.
Dark matter of this kind can behave as if there is no dark matter but
MOND.  Therefore, we call it ``MONDian dark matter"  \cite{HMN}.
One can solve for $M'$ as a function of $r$ in the two acceleration
regimes: $M' \approx 0$ for $a \gg a_0$, and (with $a_0 \sim
\sqrt{\Lambda}$)
\begin{equation*}
M' \sim (\sqrt{\Lambda}/G)^{1/2} M^{1/2} r
\end{equation*}
for $a \ll a_0$.
Intriguingly {\it the dark matter profile} we have
obtained {\it relates, at the galactic scale},\footnote{
One may wonder why MOND works at the galactic scale, but not at
the cluster or cosmic scale.  One of reasons is that, for the larger
scales, one has to use Einstein's equations with non-negligible
contributions from the pressure and explicitly the cosmological constant,
which have not been taken into account in the MOND scheme \cite{HMN}.}
{\it dark matter ($M'$), dark energy ($\Lambda$) and ordinary matter
($M$) to one another}.  As a side remark,
this dark matter profile can be used to recover the observed flat
rotation curves and the Tully-Fisher relation.

\section{Inertia in Quantum Gravity}

As the example of MOND illustrates, the inertia -- the response of a body
to force -- is not an inherent property of bodies.  It depends on the
background medium.  Let us discuss this
phenomenon in a more general and wider context.  Adopting the
thermo-field-dynamics language of Israel \cite{Israel} and
others \cite{Mald,VanRaam},
we start from the familiar expansion of the Minkowski vacuum $\psi$
in terms of an ensemble sum over
the entangled left and right Rindler vacuum $\phi$:
$\psi_M = \sum_i e^{-\beta E_i} \phi^L_i \times \phi^R_i$,
where $\beta$ is the inverse of the Unruh temperature $T \sim \hbar a/c$.
Better yet, instead of $\phi^{L,R}$ we can use, in the above formula,
the exact quantum gravity states, such as fuzzball
states introduced by Mathur \cite{Mathur}, the presumed exact no-horizon
states of string theory (viewed as a quantum theory of gravity).

Replacing the Minkowski vacuum with the global de Sitter vacuum and
the entangled left and right Rindler patches with the entangled left and right static
patches of de Sitter (or better still, with their fuzzball counterparts)
we can expand the de Sitter vacuum with cosmological constant $\Lambda$
as
\begin{equation*}
\Psi^{\Lambda}_{dS} = \sum_i e^{-\beta_{dS} E_i} \Phi^L_i \times
\Phi^R_i.
\end{equation*}
This formula is in the spirit of the discussion of holography in de Sitter space
as presented in \cite{vijay}.
Here the Unruh temperature in the de Sitter space is
given by $\tilde{T} = [(a^2+a_0^2)^{1/2} - a_0]/(2 \pi k_B)$ with
$a_0=\sqrt{\Lambda/3}$ introduced above.  The weighting factors in the
ensemble sum depend on the energies
$E_i \sim m_i c^2$ with $m_i$ being the invariant masses associated with
the exact quantum gravity spectrum (which we do not know how to
calculate at present).
Thus, for the Minkowski case, we have ratios like $m_i a^{-1}$ in the ensemble
weights (in the units
of $c = 1, \hbar = 1$ and $k_B = 1$).
But, for the de Sitter case, we obtain these ratios only in the limit of
$a \gg a_0$, whereas in the limit of
$a \ll a_0$ the ratios are $m_i (a^2/a_0)^{-1}$, where $a_0$ is the critical
acceleration a la Milgrom.
The latter result for the de Sitter case can be reinterpreted to
mean that {\it the inertial properties of massive particles} with
mass $m_i$
{\it have become acceleration- and
$\Lambda$-dependent}!
(This is analogous to what happens to the inertial properties of
particles in special relativity, where the dependence is on
the velocity of the particle and $c$, the speed of light.)

Therefore, the weights of the exact formula which represents
the expansion of the exact de Sitter quantum gravity state
in terms of entangled exact stationary quantum gravity states, such as
fuzzballs, associated with the left and right static patches,
are scale-dependent, and the weights determine what we
mean by the mass
at different scales.  This
general argument suggests that
Milgrom's scaling is just a new physical phenomenon in which
the inertial properties are acceleration- and
$\Lambda$-dependent.

\section{Discussion}

To recapitulate, we have given an entropic and holographic dual
description of the Milgrom scaling associated with galactic rotation
curves and the Tully-Fisher relation by showing
how dark matter can emulate the modified Newtonian dynamics at
the galactic scale.
In addition we have argued that
the inertia of matter is quantum gravitational and is deeply connected to
dark energy.

The last point reminds us of the relation
between nuclear physics and the theory of quantum
chromodynamics (QCD).  We can make the following analogy:
the phenomenological lagrangians of nuclear physics would
correspond to the phenomenological treatments of
matter in particle physics.  But the true picture should be
found in quantum gravity, the analog of QCD in this comparison.
Furthermore, the Milgrom scaling can perhaps
be viewed as an analog of the Bjorken scaling
and quantum gravity corrections to the Unruh formula as
the analog of the logarithmic corrections to the Bjorken scaling
as provided by asymptotic freedom in QCD.

We conclude with some open questions to be investigated.  Our
arguments have been mainly thermodynamical, hence the precise
nature of MONDian dark matter is still unclear; we should construct
a microscopic theory which would also address the question of
crossover between the Newtonian and MOND regimes.  The dark
matter profile we have obtained (at the galactic scale) hints at a
fixed energy density ratio between the three different
cosmological components of the Universe; how does this relation
help to alleviate the coincidence problem?  That same relation
seems to suggest that the microscopic MONDian dark matter degrees of
freedom may have knowledge of the non-local cosmological constant;
what are some of the phenomenological implications (for example, in dark
matter searches)?
Because we have the exact formula for the Unruh acceleration, we
can now look at the higher order terms (in powers or inverse powers of
$a^2/\Lambda$)
in the expansion of the relevant expressions
to examine possible phenomenological consequences. For example, are
there corrections to the Milgrom scaling?  Would the corrections
improve MOND's agreement with the rotation curves?  What are the
effects on the Tully-Fisher relation?  Are the correction terms
relevant to sub-galactic scales?  Finally we have given a general
argument showing that inertia of a body depends on the physical
vacuum.  Can we derive the entropic force directly in this general
framework?  And what novel properties of dark matter and distinctive
phenomenologies can be uncovered?

\vskip 0.5cm

\noindent
{\bf Acknowledgments:}
CMH, DM, and YJN are supported in part by the US Department of Energy
under contract DE-FG05-85ER40226,
DE-FG05-92ER40677 and DE-FG02-06ER41418 respectively.

\end{document}